\documentclass[conference]{IEEEtran}
\IEEEoverridecommandlockouts
\usepackage{cite}
\usepackage{amsmath,amssymb,amsfonts}
\usepackage{algorithmic}
\usepackage{algorithm}
\usepackage{graphicx}
\usepackage{textcomp}
\usepackage{xcolor}
\usepackage{hyperref}
\usepackage{url}
\usepackage[subrefformat=parens]{subcaption}
\usepackage{siunitx}

\newcommand{\figref}[1]{Fig.\,\ref{#1}}

\newcommand{\tabref}[1]{Table\,\ref{#1}}

\newcommand{\secref}[1]{Sec.\,\ref{#1}}

\newcommand{\chk}[1]{{\color{blue}#1}}

\begin{document}

\title{A Reinforcement Learning-based Transmission Expansion Framework Considering Strategic Bidding in Electricity Markets 
}

\author{
\IEEEauthorblockN{Tomonari Kanazawa, Hikaru Hoshino, Eiko Furutani}
\IEEEauthorblockA{\textit{Department of Electrical Materials and Engineering} \\
\textit{University of Hyogo}\\
Himeji, Japan \\
\{er25q006@guh, hoshino@eng, furutani@eng\}.u-hyogo.ac.jp}
}

\maketitle

\begin{abstract}
Transmission expansion planning in electricity markets is tightly coupled with the strategic bidding behaviors of generation companies. This paper proposes a Reinforcement Learning (RL)-based co-optimization framework that simultaneously learns transmission investment decisions and generator bidding strategies within a unified training process.
Based on a multi-agent RL framework for market simulation, the proposed method newly introduces a design policy layer that jointly optimizes continuous/discrete transmission expansion decisions together with strategic bidding policies. Through iterative interaction between market clearing and investment design, the framework effectively captures their mutual influence and achieves consistent co-optimization of expansion and bidding decisions. Case studies on the IEEE 30-bus system are provided for proof-of-concept validation of the proposed co-optimization framework.
\end{abstract}


\section{Introduction}

The liberalization of electricity markets has led to increasing decentralization in investment and operation decisions among generators and transmission operators~\cite{Gomez2024}. 
Generation is controlled by private Generation Companies (GENCOs), whose decisions are driven by their aim to maximize their profits. 
Transmission development and system (and market) operation are planned by independent entities, System Operators (SOs) and Market Operators (MOs).
Under such competitive environments, transmission system operators are required to conduct \textit{proactive planning}~\cite{Sauma2006} that accounts for uncertainties in future generation and market outcomes. 
Traditional multi-level optimization models have addressed transmission expansion planning together with market operation (see, e.g.,~\cite{Sauma2007,Garces2009}), yet these models typically assume rational market equilibrium and have limited flexibility to fully represent strategic bidding behaviors and price volatility.

Recent studies have applied deep Reinforcement Learning (RL) to planning of power systems in uncertain and decentralized environments. 
As reviewed in~\cite{Pesantez2024}, existing studies have used RL in two distinct ways: operational planning and expansion planning. 
The first direction uses RL to simulate operation and market behavior, where agents representing GENCOs learn strategic bidding policies that reproduce realistic price formation and market equilibria 
(see, e.g.,~\cite{Ye2020,Liang2020,Du2021}). 
These models capture short-term interactions among agents and enable analysis of market efficiency, price volatility, and regulatory impacts. 
The second direction uses RL or related learning techniques  for expansion planning and investment, such as transmission expansion problems~\cite{MingKui2020,Wang2021,Dong2025}. 
In these studies, RL serves as a high-level optimizer that explores design or investment decisions over long-term horizons, often neglecting operational details into simplified cost/reward functions.
However, these two research lines have largely developed in parallel. 
Operational optimization or market-simulation approaches using RL provide realistic operational dynamics but typically assume fixed network configurations, 
while RL-based expansion planning approaches ignore how strategic bidding behaviors affect investment outcomes. 
In practice, transmission expansion planning is inherently coupled with market behavior: network upgrades influence locational prices and dispatch patterns, 
and conversely, bidding strategies and congestion patterns determine the economic value of transmission expansions.

To bridge this research gap, this paper proposes an RL-based co-optimization framework
that incorporates strategic bidding and transmission expansion within a unified learning process. 
Based on the multi-agent deep RL framework for market simulations as reported in~\cite{Du2021}, we extend it to jointly optimize transmission investment decision as design variables that co-evolve with agent policies. 
This type of RL-based co-optimization of system design and its operational policy has been proposed primariry in the robotics community (see, e.g.,~\cite{Schaff2019,Chen2021}) and recently applied to energy systems applications~\cite{Cauz2024,Mantani2025:PESGM}. 
To the best of our knowledge, this paper is the first to propose an RL-based co-optimization framework in the context of transmission expansion planning. 
Through case studies on the IEEE 30-bus system, we demonstrate that the proposed framework captures the interaction between market dynamics and investment decisions, yielding cost-efficient expansion plans that are consistent with strategic market behavior.

The rest of this paper is organized as follows. 
In \secref{sec:formulation}, the formulation of the transmission expansion planning problem studied in this paper is introduced. 
In \secref{sec:algorithm}, the proposed RL-based co-optimization framework is presented. 
In \secref{sec:simulation}, numerical studies on the IEEE 30-bus system are described followed by conclusion in \secref{sec:conclusion}.

\section{Problem Formulation}\label{sec:formulation}

This study considers a two-level decision process that includes short-term market operation and long-term transmission investment. The formulation below defines the underlying market model and investment variables that form the basis of the proposed RL-based co-optimization framework.

\subsection{Market Model} \label{sec:market_model}

We consider a day-ahead electricity market consisting of $N_g$ generating units (agents) participating through strategic bids.
Each agent $i$ aims to maximize its own profit by deciding a bid price $\lambda^\mathrm{bid}_i(t)$ at each time step $t$ based on observed states such as local demand and past market outcomes (the detailed modeling of bidding behavior is described in \secref{sec:rl_bidding}).
Given all submitted bids, the market operator performs DC Optimal Power Flow (DC-OPF) to clear the market under power balance and line-flow constraints at each time step $t$:
\begin{align}
    \min_{\{P_i,\theta_n\}}~ & 
        \sum_{i=1}^{N_g} \lambda_i^{\mathrm{bid}}(t) P_i(t) \label{eq:dcpopf_obj}\\
    \text{s.t. }~ 
    & \sum_{i\in \mathcal{G}_n} P_i(t) - D_n(t) \notag \\&~~=
      \sum_{(n,m)\in\mathcal{L}} B_{nm}(\theta_n(t)-\theta_m(t)), 
      ~~\forall n\in\mathcal{N}_\mathrm{b}, \label{eq:dcpopf_balance}\\
    & |B_{nm}(\theta_n(t)-\theta_m(t))| \le L_{nm},~~
      \forall (n,m)\in\mathcal{L}, \label{eq:dcpopf_flow} \\
    & 0 \le P_i(t) \le P_i^{\max},~~\forall i\in\{1,\ldots,N_g\}, \label{eq:dcpopf_genlimit}
\end{align}
where $\mathcal{N}_\mathrm{b}$, $\mathcal{L}$, and $\mathcal{G}_n$ are the sets of nodes (buses), lines, and generators connected to bus $n \in \mathcal{N}_\mathrm{b}$, respectively; 
$D_n$ is the demand at bus $n$; 
$\theta_n$ is the voltage angle at bus $n$; 
$B_{nm}$ is the line susceptance; 
$L_{nm}$ is the transmission capacity between buses $n$ and $m$; 
and $P_i^{\max}$ is the maximum generation capacity of unit $i$.  
Solving \eqref{eq:dcpopf_obj}–\eqref{eq:dcpopf_genlimit} yields 
the cleared generation $P_i^{\mathrm{cleared}}(t)$ and nodal prices $\lambda_i^{\mathrm{cleared}}(t)$, which reflect network congestion and marginal operating costs.  
The total operational system cost at time $t$ is then defined as
\begin{equation}
    C_{\mathrm{oper}}(t) = \sum_{i=1}^{N_g} 
        \lambda_i^{\mathrm{cleared}}(t) P_{i}^{\mathrm{cleared}}(t),
\end{equation}
representing the total payment to generators in the cleared market.
Although a single-block bidding structure is assumed here for simplicity, the formulation can be readily extended to multi-block or piecewise-linear bidding schemes, as commonly adopted in day-ahead markets.

\subsection{Transmission Expansion}

In this study, transmission expansion planning is considered for a single target year rather than a multi-stage horizon. 
That is, investment decisions are evaluated under representative operating conditions of a given year, and the resulting network configuration is assumed to remain fixed during the market simulations. 
Let each transmission line $\ell \in \mathcal{L}$ have a base capacity $L_{\ell}^{\mathrm{base}}$, while in \secref{sec:market_model} the pair $(n,m)$ refers to the sending and receiving buses of each line.
We consider two types of formulations of a continuous expansion model that adjusts line capacities and a discrete siting model that determines whether to upgrade candidate lines: 
\begin{align}
    \Delta L_{\ell} &\in \mathbb{R}_{+}, \quad &&\text{(continuous capacity expansion)}, \\
    z_{\ell} &\in \{0, 1\}, \quad &&\text{(discrete expansion decision)}. 
\end{align}
The effective line capacity is then expressed as
\begin{equation}
    L_{\ell} = L_{\ell}^{\mathrm{base}} + z_{\ell}\,\Delta L_{\ell},
\end{equation}
where either $z_{\ell}$ or $\Delta L_{\ell}$ is fixed depending on the case study in \secref{sec:simulation}. 
The investment cost for transmission expansion is modeled as:
\begin{equation}
    C_{\mathrm{exp}} = \sum_{\ell \in \mathcal{L}} c_{\ell} z_{\ell} \Delta L_{\ell} ,
\end{equation}
where $c_{\ell}$ is the expantion cost per unit capacity of the line $\ell$ per year. 
The overall optimization objective is to minimize the total cost including the expansion cost, defined as
\begin{equation}
    J =  W_\mathrm{anu}\sum_{t=1}^{T} C_{\mathrm{oper}}(t) + C_{\mathrm{exp}}, 
    \label{eq:total_cost}
\end{equation}
where $W_\mathrm{anu}$ stands for the annualization factor depending on the total time steps $T$.  
Such formulations are consistent with those employed in open-source capacity expansion models such as PyPSA~\cite{Brown2018PYPSA} and other optimization frameworks.

\section{RL-based Co-optimization Framework} \label{sec:algorithm}

\begin{figure}[b!]
    \centering
    \includegraphics[width=0.95\linewidth]{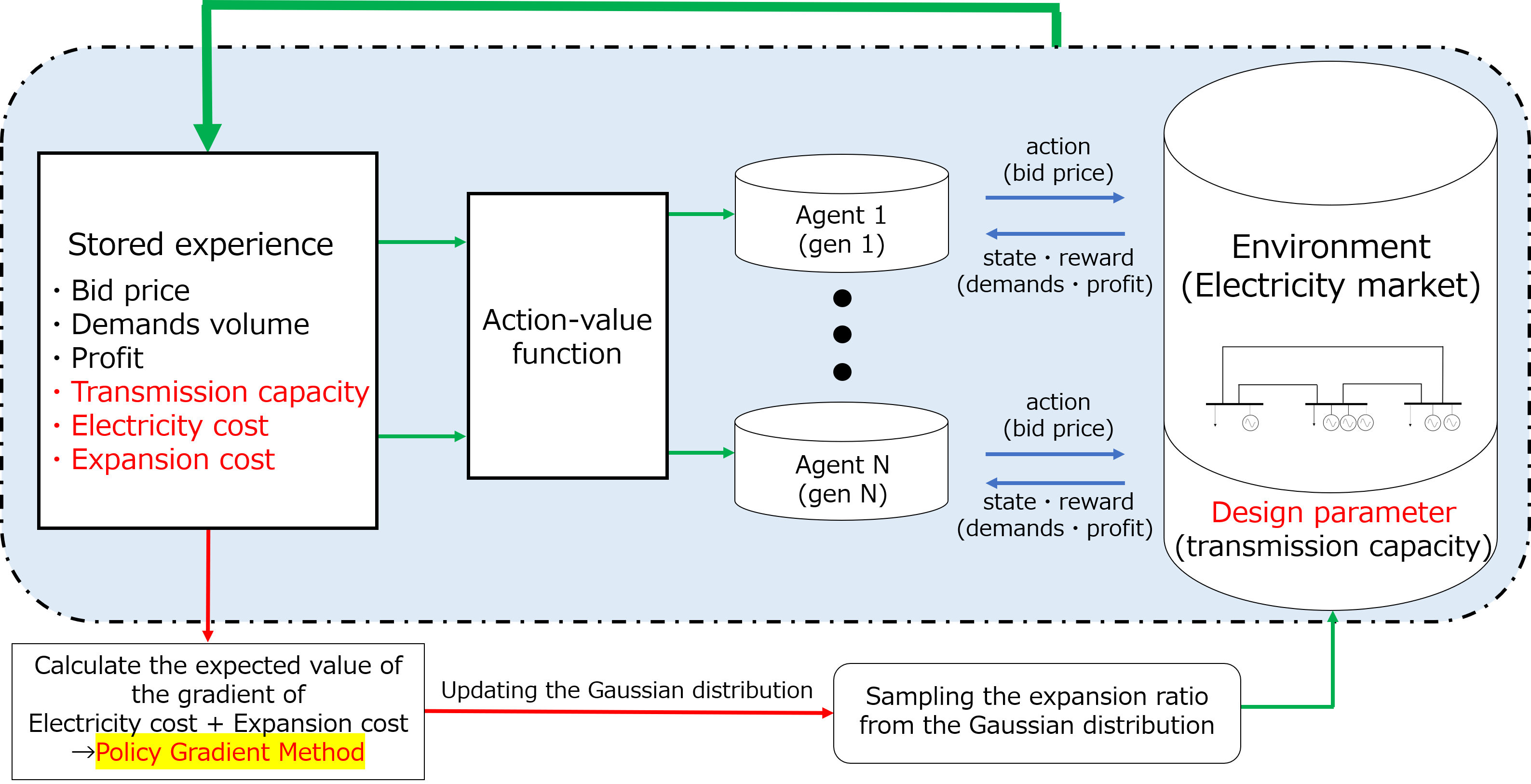}
    \caption{Overview of the proposed co-optimization framework} \label{fig:framework}
\end{figure}

The overview of the proposed co-optimization framework is shown in \figref{fig:framework}. 
The upper part of the figure (highlighted in blue) represents the multi-agent RL environment used for market simulation and is described in~\secref{sec:rl_bidding}. 
The lower part illustrates the extension that enables simultaneous optimization of transmission capacity as explained in \secref{sec:proposed_method}.  


\subsection{Market Simulation Using Multi-agent RL} \label{sec:rl_bidding}

The competitive electricity market is modeled as a partially observable multi-agent environment largely based on~\cite{Du2021}. 
Each GENCO, with a single generating unit, acts as an autonomous agent submitting strategic bids to maximize its profit.
At each time step $t$, agent $i$ observes a local state given by 
\begin{align}
 o_i(t) = [\, P_\mathrm{load}(t),\, \lambda_i^\mathrm{bid}(t-1), \, \omega^\top \,]^\top,
\end{align}
where $P_{\mathrm{load}}(t):= \sum_{n\in \mathcal{N}_\mathrm{b}} D_n(t)$ stands for the total system demand, and $\lambda_i^\mathrm{bid}(t-1)$ is the bid submitted at the previous step. 
The vector~$\omega$ represents the set of transmission design parameters that remain fixed during each episode, and its role  in the co-optimization is detailed in~\secref{sec:proposed_method}.
Based on the observation $o_i(t)$, the agent $i$ selects a normalized action $a_i(t) \in (0, 1)$, and the bid price $\lambda_i^{\mathrm{bid}}(t)$ is calculated as 
\begin{align}
    \lambda_i^{\mathrm{bid}}(t) = \lambda_i^\mathrm{cost}(t) (\alpha_i  a_i(t)+1) 
    \label{eq:bidding_price}
\end{align}
where $\lambda_i^\mathrm{cost}$ is the marginal generation cost of unit~$i$, and $\alpha_i>0$ is a scaling parameter that defines the upper limit of $\lambda_i^{\mathrm{bid}}$. 
When $a_i(t)=0$, the agent bids truthfully at its marginal cost $\lambda_i^{\mathrm{bid}}(t)=c_i(t)$, whereas larger $a_i(t)$ corresponds to higher bids. 
To avoid unrealistic temporal fluctuations in bidding, the bid prices are further constrained following~\cite{Du2021} as
\begin{align}
    \frac{\max_t \lambda_i^{\mathrm{bid}}(t)}{\min_t \lambda_i^{\mathrm{bid}}(t)} &\le 1.5, 
    \label{eq:bid_limit_day}\\
    0.9 \le \frac{\lambda_i^{\mathrm{bid}}(t)}{\lambda_i^{\mathrm{bid}}(t-1)} &\le 1.1. 
    \label{eq:bid_limit_hour}
\end{align}
After all agents submit their bids, the market operator clears the market as described in~\secref{sec:market_model}, and each agent receives a reward $r_i$ calculated as 
\begin{equation}
    r_i(t)
    = (\lambda_i^{\mathrm{cleared}}(t)-  \lambda_i^\mathrm{cost}(t)) P_i^{\mathrm{cleared}}(t),
\end{equation}
and updates its policy to maximize the long-term return
\begin{equation}
    G_i = \sum_t \gamma^t r_i(t), \label{eq:return}
\end{equation}
where $\gamma \in (0,1)$ is the discount factor.

To address the coupled and non-stationary nature of multi-agent bidding, we
employ the Multi-Agent Deep Deterministic Policy Gradient
(MADDPG) algorithm.  
Each agent possesses an actor network $\pi_i(o_i;\theta_i^{\pi})$ that maps local observations to continuous bid
actions, and a centralized critic
$Q_i(\mathbf{o},\mathbf{a};\theta_i^{Q})$
that evaluates the joint action--value function using all agents'
observations $\mathbf{o}=(o_1,\dots,o_{N_g})$ and actions
$\mathbf{a}=(a_1,\dots,a_{N_g})$.
During training, the centralized critic mitigates non-stationarity,
while each actor learns its decentralized policy for execution.
After each market clearing, experience tuples
$(\mathbf{o}_t,\mathbf{a}_t,r_t,\mathbf{o}_{t+1})$
are stored in a replay buffer and used for off-policy updates of both actor and
critic networks.
This procedure allows the agents to learn Nash-like bidding equilibria
consistent with network congestion and market clearing.

\subsection{Co-optimization of Transmission Capacity} \label{sec:proposed_method}

To enable co-optimization of transmission expansion, we treat the design vector $\omega = [\omega_1, \dots, \omega_{|\mathcal{L}|}]^\top$ as a random variable sampled at the beginning of each episode. 
Its sampling follows a design policy parametrized by $\mu=\{\mu_\ell\}_{\ell \in \mathcal{L}}$: 
\begin{align}
p_{\mu}(\omega)=\prod_{\ell\in\mathcal{L}}p_{\mu_\ell}(\omega_\ell), 
\end{align}
where $\mu_\ell$ is the learnable parameter governing the distribution of the design variable for line~$\ell$.
In the continuous case, $\omega_\ell$ corresponds to the capacity increment $\Delta L_\ell$, and we use a Gaussian distribution $\mathcal{N}(\mu_\ell,\sigma_\ell^2)$ as the design policy:
\begin{align}
p_{\mu_\ell}(\omega_\ell)
= \frac{1}{\sqrt{2\pi\sigma_\ell^{2}}}
\exp\left[-\frac{(\omega_\ell-\mu_\ell)^{2}}{2\sigma_\ell^{2}}\right],
\end{align}
where $\sigma_\ell$ is a fixed standard deviation controlling exploration.
In the discrete case, $\omega_\ell$ corresponds to the binary line-upgrade decision $z_\ell\in\{0,1\}$, for which we use a Bernoulli distribution:
\begin{align}
    p_{\mu_\ell}(\omega_\ell) = \mu_\ell \delta(\omega_\ell-1) +(1-\mu_\ell)\delta(\omega_\ell),
\end{align}
where the probability density is expressed using Dirac delta function $\delta(\cdot)$ for notational consistency, even though a discrete random variable is normally described by a probability mass function.
The objective of the design-policy update is to minimize the total cost, and thus the episodic return $G_\mathrm{total}$ is defined as
\begin{align}
G_\mathrm{total}
= -\Big(
W_{\mathrm{anu}}
\sum_{t=1}^{T} C_{\mathrm{oper}}(t)
+ C_{\mathrm{exp}}
\Big).
\end{align}
Based on this objective, the design-policy parameter 
$\mu$ is updated using the policy-gradient theorem~\cite{Sutton2018:RL}:
\begin{align}
\nabla_{\mu}\mathbb{E}[G_\mathrm{total}]
= \mathbb{E}\left[\nabla_{\mu}\ln p_{\mu}(\omega) (G_\mathrm{total} - \bar{G}_\mathrm{total}) \right],
\end{align}
where $\nabla_{\mu}$ is the gradient operator with respect to
$\mu$, and $\bar{G}_\mathrm{total}$ is a moving-average of $G_\mathrm{total}$ introduced to reduce the variance of the gradient estimator.
The expectation is approximated by the sample average over the most recent $N_\mathrm{up}$ episode, and $p_{\mu}$ can be replaced with a probability mass function for discrete variables. 
The overall learning flow 
is summarized in Algorithm~\ref{alg:co-optimization}.

\begin{algorithm}[t!] 
\caption{Co-optimization of Bidding and Expansion} \label{alg:co-optimization}
\begin{algorithmic}[1]
\STATE Initialize $\theta^{\pi}_i,\theta^{Q}_i$ and parameter $\mu$
\FOR{episode $=1$ to $N$}
  \STATE Sample design $\omega\!\sim\!p_\mu(\omega)$
  \STATE Simulate market episode with MADDPG bidding
  \STATE Compute episodic returns $G_i$
  \STATE Update actor and critic parameters $\theta^{\pi}_i,\theta^{Q}_i$
  \IF{episode $\bmod N_{\mathrm{up}}=0$}
    \STATE Update $\mu \gets \mu +\alpha \nabla_\mu\mathbb{E}[G_\mathrm{total}]$
  \ENDIF
\ENDFOR
\end{algorithmic}
\end{algorithm}

\section{Numerical Experiments}
\label{sec:simulation}

This section evaluates the proposed framework through proof-of-concept examples on the IEEE 30-bus system.
After presenting the system settings in \secref{sec:settings}, two types of experiments are conducted.
In \secref{sec:continuous}, we compare the proposed method with a conventional two-stage benchmark using PyPSA~\cite{Brown2018PYPSA} under continuous expansion for two selected lines.
In \secref{sec:discrete}, we examine a more realistic setup involving discrete siting decisions over multiple candidate lines.

\subsection{System Settings} \label{sec:settings}

\begin{figure}[!t]
    \centering
    \includegraphics[width=0.7\linewidth]{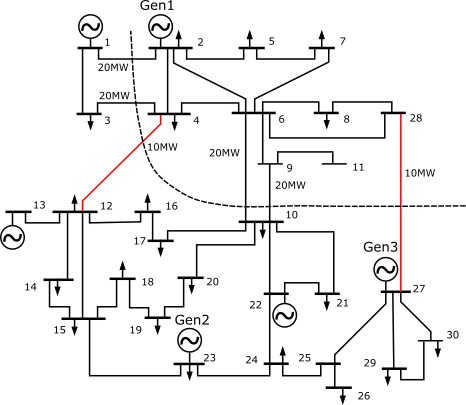}
    \caption{Topology of IEEE 30-bus system} \label{fig:30bus_system}
    \vspace{-3mm}
\end{figure}

The topology of the IEEE 30-bus system is shown in \figref{fig:30bus_system}. 
The generators at bus 2 (Gen~1), bus 23 (Gen~2), and bus 27 (Gen~3) are considered as strategic bidders that maximize their profits, and all other generators submit their true marginal cost. 
The marginal generation cost is assumed constant over the output range, and is set to $\$50\mathrm{/MWh}$ for strategic bidders and $\$55\mathrm{/MWh}$ for other generators. 
In this system, a relatively large demand is concentrated in the upper-right area, and the lines connecting to this region are selected as candidates for expansion.
The baseline transmission capacities of lines 1-2, 3-4, 6-10, and 9-10 are set to 20 MW, while the capacities of lines 4-12 and 27-28 are set to 10 MW. The capacity expansion cost is assumed to be \$100,000/MW/year.
The market simulations are performed with an hourly resolution over $T = 48$ intervals for illustration, though longer horizons are readily scalable within the RL framework.

The bidding agents are trained with MADDPG, where each actor network consists of six fully connected layers and each critic network consists of four layers. 
All hidden layers has 128 units and use ReLU activation.
Adam learning rates are set to $10^{-7}$ (actor) and $10^{-5}$ (critic), with discount $\gamma=0.99$, target-network smoothing $\tau=5\times10^{-3}$, replay buffer size $2\times10^{4}$, and minibatch size 64. 
Actions are bounded in $[0,1]$, and the parameter $\alpha_i$ in \eqref{eq:bidding_price} is set to~1.

\subsection{Case Study 1: Comparison with PyPSA Benchmark} \label{sec:continuous}

This subsection compares the proposed RL-based co-optimization with a benchmark using PyPSA as a standard capacity expansion model.
Continuous capacity expansion is considered for two selected lines (4-12 and 27-28).
Since PyPSA does not account for strategic bidding in its transmission planning, a two-stage benchmark is considered to emulate the conventional sequential process.
Specifically, we compare:
\begin{itemize}
  \item \textbf{Two-stage benchmark }:
  (Stage1) Use PyPSA to compute continuous capacity expansions $\Delta L_\ell$ under \emph{exogenous} bidding assumptions (e.g., “high-submission” and “low-submission” regimes);
  (Stage2) Given those capacities, simulate the multi-agent market with RL to obtain the realized operational outcomes.
  \item \textbf{Proposed RL-based co-optimization}:
  Learn bidding policies and continuous expansion $\Delta L_\ell$ simultaneously within the unified RL framework described in \secref{sec:algorithm}. 
  \end{itemize}

\begin{figure}[!t]
    \centering
    \includegraphics[width=0.95\linewidth]{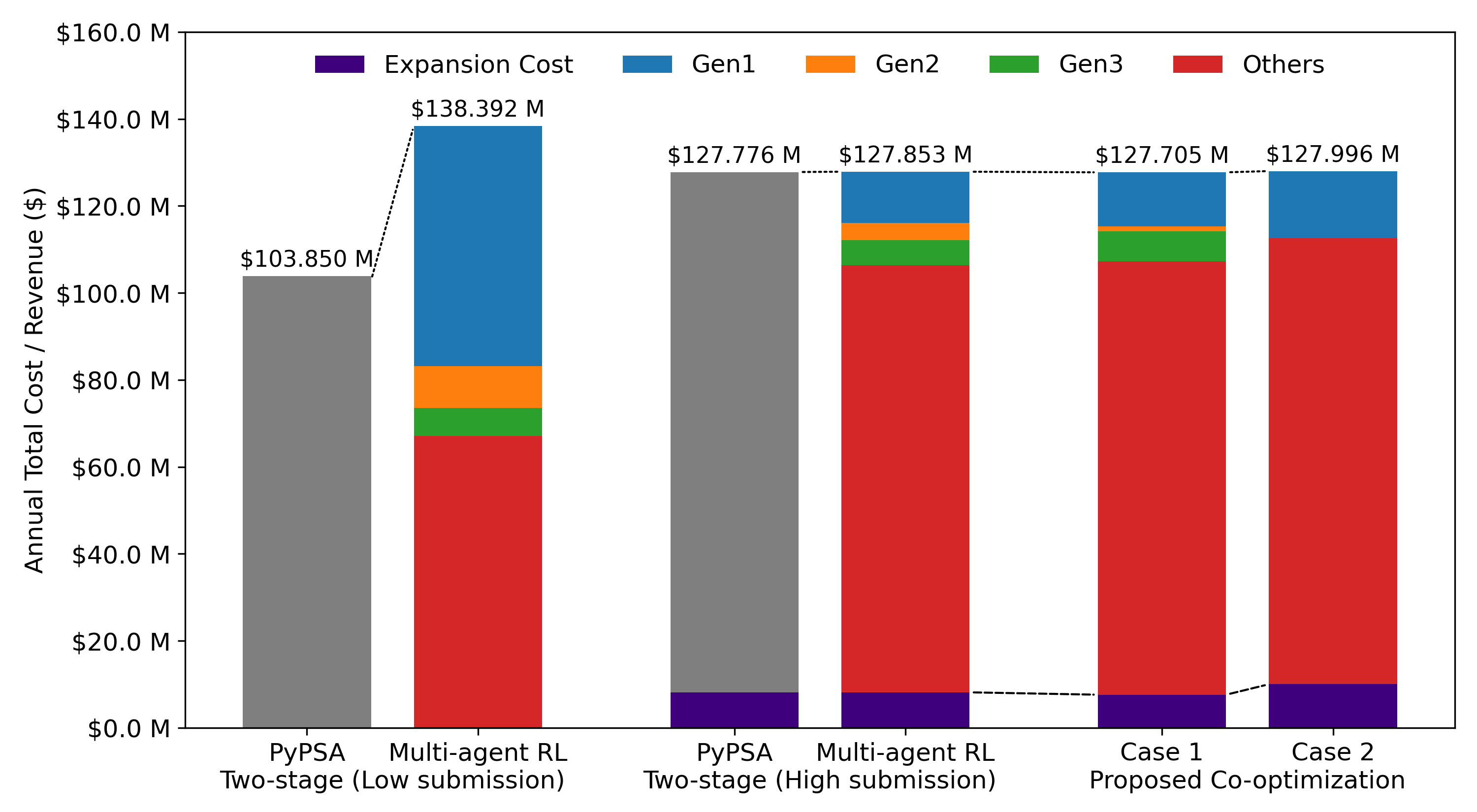}
    \caption{Total cost comparison and component breakdown}
    \label{fig:total_cost}
\end{figure}

We first present the results of the two-stage benchmark.
The low-submission scenario assumes that all strategic generators bid at their marginal cost of $\$50\mathrm{/MWh}$, while the high-submission scenario assumes more aggressive bidding at $\$90\mathrm{/MWh}$ determined based on preliminary RL simulations.
The former leads to no expansion in PyPSA, while the latter installs $\SI{81.4}{MW}$ 
on Line 4-12, with Line 27-28 unchanged.
The resulting market outcomes obtained by running multi-agent RL under these capacities are summarized in \figref{fig:total_cost} and \tabref{tab:market_summary}\footnote{Due to stochastic exploration in RL training, the market outcomes exhibit variance across runs. For fair comparison, the figure and table report a representative run in which all three strategic generators successfully converged to stable bidding policies, ensuring that the underlying learning dynamics are consistent across the evaluated cases.}.
The gray and purple bars in \figref{fig:total_cost}  show the total \emph{operational cost} and \emph{expansion cost} given by PyPSA in Stage~1, while the stacked colored bars indicate the realized generator \emph{revenues} (rather than \emph{profit}) under the subsequent market simulation.
Under the low-submission assumption, the lack of expansion results in severe congestion, resulting in a substantially higher realized total cost compared with the PyPSA estimate.
Under the high-submission assumption, the expansion on Line~4-12 greatly reduces congestion, and the realized cost remains close to the PyPSA estimate. 
A slight increase is observed because, in PyPSA, the \$90/MWh bids of Gen~2 and Gen~3 prevent them from being dispatched at all, whereas in the RL simulation they sometimes bid above the \$55/MWh marginal cost of the non-strategic generators (as shown in \tabref{tab:market_summary}) and are dispatched at these higher prices, increasing the realized total cost.

\begin{table*}[t]
  \centering
  \renewcommand{\arraystretch}{1.05} 
  \caption{Summary of market simulation outcomes in case studies}
  \label{tab:market_summary}
  \begin{tabular}{c|l|ccc|ccc|ccc}
    \hline
    Case & Scenario 
         & \multicolumn{3}{c|}{Average bid price [\$/MWh]} 
         & \multicolumn{3}{c|}{Profit [M\$]} 
         & Operational cost
         & Expansion cost \\
    \cline{3-5} \cline{6-8} & 
         & Gen~1 & Gen~2 & Gen~3
         & Gen~1 & Gen~2 & Gen~3
         & [M\$]
         & [M\$] \\
    \hline
      & Two-stage (low-submission) 
      &  93.7
      &  58.1
      &  69.8
      &  26.2
      &  0.66
      &  1.62
      &  138.39
      &  0
      \\
    1 & Two-stage (high-submission) 
      & 93.0
      & 58.0
      & 63.7
      & 5.42
      & 0.274
      & 1.24
      & 119.71
      & 8.14
      \\
      & Proposed co-optimization (continuous) 
      & 91.5
      & 60.1
      & 62.4
      & 5.72
      & 0.483
      & 1.43
      & 120.09
      & 7.61
      \\
    \hline
    2 & Proposed co-optimization (discrete) 
      & 75.6
      & 94.2
      & 84.6
      & 4.82
      & 0
      & 0.0351
      & 118.00
      & 10.00
      \\
    \hline
  \end{tabular}
\end{table*}

In contrast to the two-stage benchmark, the proposed RL-based co-optimization jointly learns both the bidding strategies and the transmission expansion.
The learned design results in an expansion of $\SI{76.1}{MW}$ on Line~4-12. 
Compared with the two-stage benchmark, particularly the high-submission case, the co-optimized solution achieves a slight reduction in expansion cost. 
The two-stage approach assumes uniformly high bids for all strategic generators and therefore installs more capacity than necessary, whereas the proposed method internalizes the strategic behaviors and selects a more efficient capacity level.
While such integrated planning could in principle be formulated using multi-level optimization such as~\cite{Sauma2007,Garces2009}, the multi-agent RL-based approach offers 
superior scalability to longer simulated time horizons
and can readily accommodate more complex settings such as multi-market environments.

\chk{

}

\subsection{Case Study: Discrete Siting Decisions} \label{sec:discrete}

To demonstrate the applicability of the proposed framework to more realistic planning problems, we next consider a discrete siting case.
The six candidate lines introduced in \secref{sec:settings} are included as binary expansion decisions, and the additional capacity is fixed at $\SI{50}{MW}$ for all lines.
Under this setting, the design policy learns to upgrade two candidate lines 1–2 and 4-12. 
Although the expansion cost increases relative to the results in case~1 (\secref{sec:continuous}),  due to the additional upgrade of Line~1-2, the operational cost decreases sufficiently to yield a total cost that remains comparable to Case~1.
A key factor behind this operational cost reduction is the change in Gen~1’s bidding behavior induced by the upgrade of line~1-2.
In Case~1, even with the expansion of line~4-12, Gen~1 still enjoyed significant market power, and thus maintained high bid prices around \$90/MWh.
By contrast, the additional capacity on line~1-2 reduces congestion on this corridor Gen~1’s ability to exert such market power.
These results demonstrate that the proposed framework can identify effective expansion sites by internalizing the impact of strategic market behavior.


\section{Conclusions}
\label{sec:conclusion}

This paper presented an RL-based framework for co-optimization of strategic bidding and transmission expansion planning in electricity markets. By extending a multi-agent RL algorithm with a learnable continuous/discrete design policy, the proposed method enables simultaneous learning of bidding strategies and investment decisions while capturing their mutual influence. 
The effectiveness of the proposed framework has been shown based on case studies on the IEEE 30-bus system. The proposed framework provides a promising foundation for future applications to larger systems, multi-market environments, and integrated generation-transmission co-planning.


\bibliographystyle{ieeetr}
\bibliography{ref}

@inproceedings{Mantani2025:PESGM,
    author = {T. Mantani and H. Hoshino and T. Kanazawa and E. Furutani},
    title = {Sizing of battery considering renewable energy
bidding strategy with reinforcement learning},
    booktitle = {2025 IEEE Power \& Energy Society General Meeting},
    year = {2025},
}

@article{Dong2025,
title = {Transmission expansion planning: A deep learning approach},
journal = {Sustainable Energy, Grids and Networks},
volume = {41},
pages = {101585},
year = {2025},
doi = {https://doi.org/10.1016/j.segan.2024.101585},
author = {Jizhe Dong and Jianshe Cao and Yu Lu and Yuexin Zhang and Jiulong Li and Chongshan Xu and Danchen Zheng and Shunjie Han},
}

@inproceedings{Cauz2024,
title={Reinforcement Learning for Efficient Design and Control Co-optimisation of Energy Systems},
author={Marine Cauz and Adrien Bolland and Christophe Ballif and Nicolas Wyrsch},
booktitle={ICML 2024 AI for Science Workshop},
year={2024},
pages={68},
url={https://openreview.net/forum?id=17tZF3ibk4}
}

@article{Gomez2024,
title = {Coordination of generation and transmission expansion planning in a liberalized electricity context---coordination schemes, risk management, and modelling strategies: A review},
journal = {Sustainable Energy Technologies and Assessments},
volume = {64},
pages = {103731},
year = {2024},
doi = {https://doi.org/10.1016/j.seta.2024.103731},
author = {Stefanía Gómez and Luis Olmos},
}

@Article{Pesantez2024,
AUTHOR = {Pes\'{a}ntez, Gabriel and Guam\'{a}n, Wilian and C\'{o}rdova, Jos\'{e} and Torres, Miguel and Benalcazar, Pablo},
TITLE = {Reinforcement Learning for Efficient Power Systems Planning: A Review of Operational and Expansion Strategies},
JOURNAL = {Energies},
VOLUME = {17},
YEAR = {2024},
NUMBER = {9},
ARTICLE-NUMBER = {2167},
DOI = {10.3390/en17092167}
}

@ARTICLE{Du2021,
  author={Du, Yan and Li, Fangxing and Zandi, Helia and Xue, Yaosuo},
  journal={Journal of Modern Power Systems and Clean Energy}, 
  title={Approximating Nash Equilibrium in Day-ahead Electricity Market Bidding with Multi-agent Deep Reinforcement Learning}, 
  year={2021},
  volume={9},
  number={3},
  pages={534-544},
  doi={10.35833/MPCE.2020.000502}}

@Article{Wang2021,
AUTHOR = {Wang, Yuhong and Chen, Lei and Zhou, Hong and Zhou, Xu and Zheng, Zongsheng and Zeng, Qi and Jiang, Li and Lu, Liang},
TITLE = {Flexible Transmission Network Expansion Planning Based on DQN Algorithm},
JOURNAL = {Energies},
VOLUME = {14},
YEAR = {2021},
NUMBER = {7},
ARTICLE-NUMBER = {1944},
URL = {https://www.mdpi.com/1996-1073/14/7/1944},
DOI = {10.3390/en14071944}
}

@InProceedings{Chen2021,
  title = 	 {Hardware as Policy: Mechanical and Computational Co-Optimization using Deep Reinforcement Learning},
  author =       {Chen, Tianjian and He, Zhanpeng and Ciocarlie, Matei},
  booktitle = {2020 Conference on Robot Learning},
  pages = 	 {1158--1173},
  year = 	 {2021},
  editor_ = 	 {Kober, Jens and Ramos, Fabio and Tomlin, Claire},
  volume_ = 	 {155},
  series_ = 	 {Proceedings of Machine Learning Research},
  month_ = 	 {16--18 Nov},
  publisher_ =    {PMLR},
}

@ARTICLE{Ye2020,
  author={Ye, Yujian and Qiu, Dawei and Sun, Mingyang and Papadaskalopoulos, Dimitrios and Strbac, Goran},
  journal={IEEE Transactions on Smart Grid}, 
  title={Deep Reinforcement Learning for Strategic Bidding in Electricity Markets}, 
  year={2020},
  volume={11},
  number={2},
  pages={1343--1355},
  doi={10.1109/TSG.2019.2936142}}

@ARTICLE{Liang2020,
  author={Liang, Yanchang and Guo, Chunlin and Ding, Zhaohao and Hua, Huichun},
  journal={IEEE Transactions on Power Systems}, 
  title={Agent-Based Modeling in Electricity Market Using Deep Deterministic Policy Gradient Algorithm}, 
  year={2020},
  volume={35},
  number={6},
  pages={4180--4192},
  doi={10.1109/TPWRS.2020.2999536}}

@INPROCEEDINGS{MingKui2020,
  author={MingKui, Wei and ShaoRong, Cai and Quan, Zhou and Xu, Zhou and Hong, Zhou and YuHong, Wang},
  booktitle={2020 IEEE Sustainable Power and Energy Conference (iSPEC)}, 
  title={Multi-objective transmission network expansion planning based on Reinforcement Learning}, 
  year={2020},
  volume={},
  number={},
  pages={2348--2353},
  doi={10.1109/iSPEC50848.2020.9350990}}

@INPROCEEDINGS{Schaff2019,
  author={Schaff, Charles and Yunis, David and Chakrabarti, Ayan and Walter, Matthew R.},
  booktitle={2019 International Conference on Robotics and Automation (ICRA)}, 
  title={Jointly Learning to Construct and Control Agents using Deep Reinforcement Learning}, 
  year={2019},
  pages={9798-9805},
  doi={10.1109/ICRA.2019.8793537}}

@book{Sutton2018:RL,
author  = {Richard S. Suttton and Andrew G. Barto},
title   = {Reinforcement Learning: An Introduction},
edition = {2nd},
year    = {2018},
publisher  = {MIT Press},
}

@article{Brown2018PYPSA,
  title={{PyPSA}: Python for Power System Analysis},
  author={Brown, Thomas and H{\"o}rsch, Jonas and Schlachtberger, David},
  journal={Journal of Open Research Software},
  volume={6},
  number={1},
  year={2018}
}

@ARTICLE{Garces2009,
  author={Garces, Lina P. and Conejo, Antonio J. and Garcia-Bertrand, Raquel and Romero, RubÉn},
  journal={IEEE Transactions on Power Systems}, 
  title={A Bilevel Approach to Transmission Expansion Planning Within a Market Environment}, 
  year={2009},
  volume={24},
  number={3},
  pages={1513--1522},
  doi={10.1109/TPWRS.2009.2021230}}

@ARTICLE{Sauma2007,
  author={Sauma, Enzo E. and Oren, Shmuel S.},
  journal={IEEE Transactions on Power Systems}, 
  title={Economic Criteria for Planning Transmission Investment in Restructured Electricity Markets}, 
  year={2007},
  volume={22},
  number={4},
  pages={1394--1405},
  doi={10.1109/TPWRS.2007.907149}}

@article{Sauma2006,
  title={Proactive planning and valuation of transmission investments in restructured electricity markets},
  author={Sauma, Enzo E and Oren, Shmuel S},
  journal={Journal of Regulatory Economics},
  volume={30},
  number={3},
  pages={358--387},
  year={2006},
  publisher={Springer}
}


\end{document}